# A complete demonstrator of a muon cooled Higgs factory.

Carlo Rubbia

*GSSI, L'Aquila, Italy*

*Institute for Advanced Sustainability Studies Potsdam, Germany*

**Abstract**

The recent discovery of the Higgs particle of 125 GeV has revised the interest of the so-called Higgs factory, namely of a $\mu^+ - \mu^-$ collider with adequately cooled intensity of about 6 x $10^{12}$ muons of each sign, a repetition rate of 15-50 p/s and a luminosity of up to $10^{32}$ cm$^{-2}$ s$^{-1}$. The process is the direct production of a H$^o$ scalar in the s-state. Its partial widths can be studied in clean conditions and with remarkable accuracy. The H$^o$ mass itself can be measured to about ±100 keV ($\Delta E/E \leq 10^{-6}$) in the (WW) channel with the help of the decay frequency of the polarized $\mu \rightarrow e\nu\nu$ decay electrons.

The realization of a cascade of unconventional but very small rings of few meters radius is described, in order to achieve the required longitudinal and transverse emittances. Physics requirements and the many studies already undertaken suggest that the next step, prior to but adequate for a H$^o$ physics programme, could be the practical realization of a full scale muon cooling demonstrator. The correct operation of the demonstrator may be initially explored with the help of very low intensity muon beam already available in a number of different accelerators.

The additional but relatively conventional components necessary to realize the facility with the appropriate luminosity should be constructed only after the success of this initial cooling experiment has been conclusively demonstrated. The ultimate $\mu^+ - \mu^-$-collider for a Higgs Factory can be situated within the existing CERN or FNAL sites.

August 16, 2013

## 1.— Introduction.

The ATLAS [1] and CMS [2] teams have observed at the CERN-LHC a narrow line of high significance at a mass of about 125 GeV, compatible with the Standard Model (SM) Higgs boson. Results of both experiments also exclude other SM Higgs bosons from 127 GeV up to approximately 600 GeV. Observations have been performed in several decay modes, however always in the presence of very substantial backgrounds.

It had been argued by many theorists that "new physics" must also necessarily appear at the TeV scale, one of the main reasons for arguing for the necessity of a nearby SUSY[3]. This was based on the argument that the otherwise divergent self-interaction of the Higgs scalar does require the presence of a cut-off at the TeV scale. However, this does not hold for the recently observed Higgs mass of 125 GeV, since now stability conditions may allow without novelties a legitimate cut-off. The Higgs potential develops instability [4] only around $10^{11}$ GeV, with a lifetime much longer than the age of the Universe. However, taking into account theoretical and experimental errors, stability up to the Planck scale cannot be excluded [4]. Therefore there is no need for the so called "no fail theorem" and there may be only one Higgs particle to be studied experimentally.

During the next twenty years LHC plans are to pursue the hadronic production of the Higgs related sector and of the possible additional existence of SUSY. The existence of additional Higgs particles is unlikely within the LHC energy range. Therefore studies should concentrate on the properties of the already discovered mass. The High Luminosity-LHC option which should reach an integrated sensitivity of $10^{42}\,\text{cm}^2$ will be already a sort of "Higgs factory", able to perform relatively accurate (typically ± 10%) measurements.

The scalar sector is definitely one of the keys to the understanding of elementary particle physics. Several other "exotic" alternatives [4] have been theoretically considered in order to conclusively confirm or disprove the validity of the SM Higgs. Sensitivity to new physics and "5 sigma" discoveries may need however per-cent to sub-per-cent accuracies, much better than the ultimate LHC expectations.

The SM Higgs boson has several substantive branching fractions which need to be accurately compared with the experiments: (bb), 60%, (WW), 20%, (gg), 9%, (ττ), 6%, (ZZ), 3%, (cc), 3%. The process (γγ) with 0.2% is also substantive due to the high mass resolution and relatively low background. In particular, like in the case of the studies on the $Z^o$, the determination of the actual $H^o$ width will be crucial for the determination of the nature of the particle and the underlying theory. However the SM prediction for the Higgs width is only ≈ 4.2 MeV (Figure 1).

In the previous example of the $Z^o$, detailed studies at LEP and SLAC in very clean conditions have been the necessary second phase after its initial discovery with p-pbar hadronic collisions. A similar procedure is doubtless required also in the case of the H⁰ with a detailed search of the decay channels in much cleaner conditions. Two main alternatives are possible and they are hereby compared:

1) A $e^+e^-$ collider with L > $10^{34}$ and with a Z+H⁰ signal of ≈ 200 fb. The beam energy is of the order of 200 GeV. The accelerator may be either a storage ring or a linear collider. The circular machine option is more conservative but it requires a circumference of about 60-80 km. (see for instance the proposals for TLEP [5] at CERN or SuperTRISTAN [6] in Japan: similar proposals have been discussed also elsewhere). The luminosity must be pushed to the beam-strahlung limit, i.e. to about 400 times [6] the highest luminosity achieved with LEP2 (L = 1.2 x$10^{32}$). The vertical emittance is extremely small, with a beam crossing size the order of 0.07 μ (it was 3.5 μ for LEP2). The beam lifetime is very short, of the order of 20 minutes. The RF power consumption is very high, in the order of hundreds of MWatt.

2) A $\mu^+ - \mu^-$ collider at L > $10^{32}$ and a H⁰ signal of ≈ 20'000 fb in the s-state [7]. The collider ring is much smaller, with R ≈ 50 m. However a powerful "muon cooling" is necessary [8]. In a $\mu^+ - \mu^-$ collider the direct H⁰ cross section is greatly enhanced with respect to $e^+ - e^-$ since the s-channel coupling to a scalar is proportional to the square of the initial lepton mass. In analogy to the case of the Z⁰, the production of a single H⁰ scalar in the s-state offers unique conditions of cleanliness. A feature of this method is that it actual mass, its very narrow width and most of its decay channels may be directly compared to the SM predictions with a very high accuracy. The properties of the Higgs boson can be detailed [7] over a larger fraction of model parameter space than at any other proposed accelerator method. A particularly important conclusion is that the $\mu^+ - \mu^-$ collider will have greater potentials for distinguishing between a standard SM and the SM-like H⁰ of SUSY or of other models than with any other alternative.

Today's experimental values of the H⁰ mass are $M_H = 125.5 \pm 0.2(stat)^{+0.5}_{-0.6}(syst)$ [1] and $M_H = 125.8 \pm 0.4(stat) \pm 0.4(syst)$ GeV [2]. No doubt more accurate determinations will be possible in the future with the LHC and a window of the order of ± 200 MeV should be reasonable in the future. The $\mu^+ - \mu^-$ collider should observe the natural H⁰ width of a few MeV. A much higher accuracy, of the order of ±100 keV, may be possible with the g-2 precession of polarized muons [20] and an adequate luminosity.

## 2.— From protons to muons.

Over the past decades, there has been significant progress in developing the concepts and technologies needed to produce, capture, cool [8][9] and accelerate as many as O($10^{13}$) muons per pulse. During the late ninety, extensive studies have been carried out in the US [10] and in several other international workshops [11] and with experiments of the MICE collaboration in the UK [12]. No doubt, the recent discov-ery of the Higgs particle at 125 GeV has strongly revived the interest for these studies [13].

Short, intense bunches of protons of several GeV and with a beam power of a few MWatt are focused onto a target to produce pions that decay into muons, which are cooled and accelerated at the relatively modest energy of 62.5 GeV, offering the possibility of $\mu^+ - \mu^-$ collisions of an adequate luminosity. The $\mu^+$ and $\mu^-$ bunches are counter-rotating in a single ring and focused to collide in two interaction regions. A tight initial proton bunch may be realized for instance with the help of an accumulation storage ring, starting from the $H^-$ beam produced by a LINAC and stripped to protons in order to produce a number of short pulses to be ultimately condensed into a single short bunch. A ≈ 5 MWatt nominal power and 10-50 cycles/s appear appropriate. The secondary pion yield is shown in Figure 2. The production rate for *a given proton power* (namely a number of protons inversely proportional to its energy) is almost independent of proton energy in the interval between 8 and 20 GeV and for 2 GeV protons it is about a factor two lower.

The resulting optimum muon momentum is in the region of 200-400 MeV/c. The best focussing is realized with secondaries in an axially symmetric focussing solenoidal field according to the so-called Bush theorem [13]. The 20-Tesla solenoidal field should collect secondary particles at an angle of about 100 mr off-axis in order to separate the intense proton beam from the secondary pions. Particles of both signs are focussed. By reducing the field the rotational motion is converted into the longitudinal one and the transverse momentum $p_t$ is reduced correspondingly. Therefore at the end of the solenoid (B ≈ 1 Tesla) we expect an average transverse momentum $<p_t> \approx$ 50 MeV/c. The MERIT/CERN experiment [14] has already successfully injected a Hg-jet into a 15-T solenoid.

Pions decay quickly into muons. After extraction from the solenoid, secondaries are magnetically separated with a pair of high field dipoles into their negative and positive components. A typical central average momentum before entering the storage rings is of the order of 200-250 MeV/c. The presently described configuration is based on a simple pair (one for each sign) of subsequent rings (Figure 7). The required initial intensity of muons of each sign is of the order of 2 x$10^{13}$ $\mu$/pulse, corresponding to a typical proton average intensity of 2.6 x $10^{14}$ p/pulse at 8 GeV and a repetition rate of 15 sec$^{-1}$.

A number of alternative methods may be considered in order to eventually further increase the number of muons and therefore the luminosity of the Higgs factory:

(1) In order to match a wider muon production energies to an acceptable $\Delta p_\mu$ spread as required in the cooling rings, a first, "Liuvillian" momentum compression can be performed with the help of a dE/dx compensating wedge equating the exiting energies of different magnetically analyzed momenta, but at the inevitable cost of a wider resulting angular spread at the entrance of the subsequent cooling ring. Positive and negative muons are then injected in separate cooling storage rings.

(2) Alternatively and in order to increase the number of accumulated muons, the cooling may be constructed as a stack of several superimposed rings (like the example of the CPS-Booster), but with different momentum slices and to be merged into a common bunch after the initial cooling.

(3) Finally it may be possible to make use of a very short proton bunch introducing with the help of appropriate RF cavities a bunch rotation of the secondary muon momentum spectrum, The RF cavity is located, after sign selection, along the muon direction channel, following an appropriate drift distance. Fast (slow) muons which travel faster (slower) arrive earlier (later) and are slowed down (accelerated). The momentum spread is reduced by the help of the RF phase, and muons are "rotated" in to a smaller momentum byte, but at the cost of an increased longitudinal extent of the secondary beam. The alternative (3) is schematically shown in Figure 3.

### 3.— The Muon cooling.

Non Liouvillian cooling is essential whenever secondary particles are produced from initial collisions and later accelerated and accumulated in a storage ring. A well known case is the one of antiprotons, in which both stochastic and electron cooling have been vastly used. As well known, P-pbar colliders have permitted the discoveries of W/Z and of the Top. At high energies, muons may be stable enough to offer a reasonable number of $\mu^+ - \mu^-$ collisions. Muon cooling is based on the ionization losses, since muons have only electromagnetic interactions with matter. The idea has been discussed in the seventies by Budker and Skrinsky [8]. A comprehensive analysis has been given in the early nineties for instance by Neuffer [9].

The method, called "dE/dx cooling" closely resembles to the damping of relativistic electrons — with the multiple energy losses in a thin, low Z absorber substituting the synchrotron radiated light. The main feature of this method is that it produces an extremely fast cooling, compared to other traditional methods. This is a necessity for the short-lived muon case. Transverse betatron oscillations are "cooled" by a target "foil" typically a fraction of g/cm$^2$ thick. An accelerating cavity is continuously replacing the lost momentum. Unfortunately for muons with $\gamma < 4$ the specific dE/dx loss is increasing with decreasing momentum. In order to "cool" also longitudinally, chromaticity has to be introduced for instance with a wedge shaped "dE/dx foil", in order to increase the ionisation losses for faster particles.

Most scenarios during the late ninety were based on single-pass linear cooler, in which a large number of RF cavities restore the energy lost in the low Z absorbers (for instance LH2 or LiH) and in the ionization cooling. However cooling rings have also been considered.

In an appropriate storage ring both transverse and longitudinal muon emittances are progressively cooled until they reach a final equilibrium state as a balance between dE/dx cooling, the RF acceleration and other effects. An analytic derivation with the description of both the transverse and longitudinal processes is given. In addition, muons spontaneously decay reducing the surviving beam fraction by a substantial amount.

The transverse emittance ε evolves toward to an equilibrium condition in which dE/dx losses are balanced by the multiple scattering (Neuffer [9] and McDonald [15])

$$\frac{d\varepsilon}{dz} \approx \frac{\varepsilon}{\beta^2 E}\frac{dE}{dz} + \frac{\beta^*(13.6)^2}{2\beta^3 E m_\mu X_o} \to 0$$

where: $z$ is the longitudinal coordinate, $dE/dz$ the ionization loss, $\beta c$ the speed of the muon of mass $m_\mu$ and total energy $E$ (in MeV), $\beta^*$ the betatron function at the dE/dx crossing point and $X_o$ the radiation length of the cooling material. The cooling process will continue until an equilibrium transverse emittance has been reached:

$$\varepsilon_N \to \frac{\beta^*(13.6\ MeV/c)^2}{2\beta_\mu m_\mu}\frac{1}{(X_o\ dE/dz)}$$

The equilibrium emittance $\varepsilon_N$ and its invariant $\varepsilon_N/\beta\gamma$ are shown in Figure 4 as a function of the stored muon momentum. For hydrogen and $\beta^*$= 10 cm and in the muon momentum interval from 80 to 300 MeV/c , $\varepsilon_N/\beta\gamma \leq 700$ mm mr. Using a solid material like for instance Li will roughly double the transverse equilibrium emittance, but it will considerably reduce the additional presence of separating windows and of the cryogenic insulation.

Longitudinal motion is related to heat producing straggling, balancing the dE/dx cooling. As already pointed out a dE/dx radial wedge is needed in order to exchange longitudinal and transverse phase-spaces. Balancing heating and cooling for a Gaussian distribution limit gives the expression [9]:

$$\frac{d(\Delta E)^2}{dz} = -2(\Delta E)^2\left[f_A\frac{d}{dE}\left(\frac{dE_o}{ds}\right) + f_A\frac{dE}{ds}\left(\frac{d\delta}{dx}\right)\frac{\eta}{E\delta}\right] + \frac{d(\Delta E)^2_{straggling}}{dz}$$

in which the first term is the intrinsic energy loss, the second is the wedge shaped absorber and the third the straggling contribution. $dE/dz = f_A\ dE/ds$ where $f_A$ is the fraction of the transport length occupied by the absorber, which has an energy absorption coefficient $dE/ds$ ; $\eta$ is the chromatic dispersion at the absorber and $\delta$ and $d\delta/dx$ are the thickness and radial tilt of the absorber. The straggling (H$_2$) is given by

$$\frac{d(\Delta E)^2_{straggling}}{dz} = \frac{\pi(m_e c^2)^2(\gamma^2 + 1)}{4\ln(287)\alpha X_o}$$

$X_o$ is the radiation length, $m_e$ the electron mass and $\alpha$ the fine structure constant. The second term gives the thickness of the wedge as an appropriate function of the transverse position. In equilibrium conditions the above indicated straggling contribution is exactly balanced by the first two terms. The equilibrium spread $(\Delta E)^2$ for a (liquid) H$_2$ absorber can then be extracted from formula 41 of Ref. [15] and it is given by

$$(\Delta E)^2 = \frac{0.55(MeV)^2 \gamma^3 \beta^4 (\gamma^2 + 1)}{(1 - \gamma^2/12)}$$

This shows (see Figure 6) a very fast dependence on the muon momentum. For instance for $p_\mu$ of 50, 85, 100, 220 and 300 MeV/c the $\Delta E_{RMS}$ varies from 0.17, 0.5, 0.74, 5 and 10 MeV. This value is in good agreement with the equilibrium value of 5 MeV of Ref. [9] for p = 230 MeV/c. Therefore for the previously indicated optimal muon producing momentum in the range of 200-300 MeV/c the final energy spread $\Delta E_{RMS}$ is exceeding the expected H$^o$ width since the SM prediction which has a predicted width is only ≈ 4 MeV (Figure 1).

The cooling process defines an equilibrium $\Delta E_{RMS}$ as a balance between cooling and straggling. The resulting longitudinal emittance $\varepsilon_L$ is a function of the relative muon momentum $\delta = \Delta p/p$ with $\Delta p_{RMS} = \Delta E_{RMS}/\beta$ and of the longitudinal half-length of the bunch $\ell_B$,

$$\varepsilon_L = \frac{1}{\pi} \int_{Area} \rho(\ell_B, \delta) d\ell_B d\delta \approx l_B \delta$$

The value $\ell_B$ depends on the actual RF used, with smaller $\ell_B$ for higher frequencies. Assuming a final $p_\mu$=100 MeV/c, $\Delta E_{RMS}$ = 0.74 MeV, a RF angle of 45° we find for $f_{RF}$ = 200 and 800 Mc/s the values $\ell_B$= 16.9 and 4.22 cm and therefore $\varepsilon_L$ = 1.72 and 0.40 mm rad respectively.

In a more realistic arrangement in which transverse and longitudinal cooling are combined, the energy cooling will also reduce somewhat the transverse cooling, according to the Robinson's law on sum of damping decrements [16].

In summary, while the transverse equilibrium distribution is within an acceptable transverse emittance (Figure 4), the longitudinal equilibrium $\Delta E_{RMS}$ can be compatible with the expected SM Higgs width only at very low muon momentum $p_\mu$, far from the optimal choice for the production yield starting from high energy protons (Figure 5). Therefore a single cooling ring at the optimal muon energy is not entirely adequate and a more sophisticated layout must be considered. A possible set-up is shown in Figure 7. After a first cooling at the optimal muon window in the region 200-300 MeV/c, the muon momentum must be substantially reduced to ensure the final momentum spread in a subsequent cooling arrangement.

### 4.— Optimal cooling for a Higgs factory.

It is then proposed a setup for both $\mu^+$ and $\mu^-$ in which a first "wide band" cooling ring at the optimal production $p_\mu \approx 220$ MeV/c introduces a first major reduction in the transverse and longitudinal emittances (but still with $\Delta E_{rms}$ > 10 MeV). The beam is then extracted and reduced by ionization losses to $p_\mu \approx$ 100 MeV/c for instance with the help of a long LH$_2$ absorber. After deceleration, the beam is finally injected in a second "deep freezer" cooling setup in order to ensure $\Delta E_{rms} \leq$ 1 MeV and the equilibrium transverse emittance $\varepsilon_N/\beta\gamma$ in order to match to the requirements of the very narrow Higgs peak.

The several components for the cooling process are now discussed. The "wide band" cooling ring (one for each muon sign) must collect the widest muon spectrum and introduce a first major reduction in the transverse and longitudinal emittances. The cooling must contain only two bunches, one for each sign. Depending on the actual configuration, it may be necessary to merge beforehand several initially produced bunches into one,

(1) Solenoids instead of quadrupoles are preferred since they have a wider acceptance, up to about ±20%. Some practical but still conceptual descriptions of RFOFO ring coolers have been given by Balbekov [17], by Palmer [18] and by Garren [19].

(2) Only a few turns are necessary; therefore only integer resonances should be considered as truly harmful.

(3) In the first "wide band" cooler, the ionization wedge absorber does not have to be made with liquid hydrogen: other solid but low Z materials (LiH) may be also used. The resulting larger emittances may be recovered later in the following cooling process.

A realistic feasibility study has been described by Garren et al. [19] and it is shown in Figure 6. The four-sided ring has four 90° arcs with 8 dipoles separated by solenoids. Arcs are achromatic both horizontally and vertically. The dispersion is zero in the straight sections between the arcs. Injection/extraction very elaborate kickers are used in a straight section and a superconducting flux pipe is used for the injected beam.

After being extracted from the ring, the deceleration of the muon momentum is performed by passing the beam through a low-Z (liquid hydrogen) absorber of the appropriate length. A one turn channel layout [18] is used in order to inject and to extract conveniently the beam. We assume that the reduced momentum after deceleration is $p_\mu$ = 100 MeV/c, corresponding to a kinetic energy $T_\mu$ = 39.7 MeV, $\beta$ = 0.686c, a muon decay length $L_{decay}$ = 622 m and $dE/dx$ = 6.86 MeV/gr cm$^2$, to be compared to the initial values $p_\mu$ = 220 MeV/c, $T_\mu$ = 138 MeV, $\beta$ = 0.90c, $L_{decay}$ = 1368 m and $dE/dx$ = 4.6 MeV/gr cm$^2$. The liquid hydrogen absorber required in order to reduce the kinetic energy from 138 to 39.7 MeV has a thickness of ≈ 2.72 m. A low β* channel and a wedge shaped Hydrogen absorber are required in order to reduce the blowup of the beam due to multiple scattering and straggling during cooldown. No RF is needed in order to correct the energy losses.

The necessary dE/dx length in the absorber is insufficient to reduce the emittances to their new equilibrium values $\Delta E_{rms} \leq 1$ MeV and $\varepsilon_N/\beta\gamma \leq 700$ mm mr (hydrogen and β*= 10 cm), Therefore an additional "deep freezer" is required. The most obvious solution that has been chosen is a so called Guggenheim helix [18] with several helical turns in which the beam is progressively passing. No kickers are needed. Since the transverse and longitudinal emittances have been already substantially cooled by the "wide band' ring, there may be no need for wide band solenoids and conventional magnets and quadrupoles can be employed. The "deep freezer" must be as compact as possible since the muon lifetime is correspondingly reduced at 100 MeV/c ($L_{decay}$ = 622 m). At this lower momentum, the variation of the ionization wedge as a function of the momentum is much more pronounced. The relative difference in the dE/dx losses for a given $\delta p/p$ difference is 4.65 times larger than the one for the "wide band" ring cooler. Fortunately the momentum spread $\delta p/p$ is also reduced. The general layout is shown in Figure 7.

The limit expected by the space-charge Laslett tune-shift is strongly dependent on the muon momentum, $\Delta Q = -3 r_\mu N_{bunch} / (2\varepsilon_o \beta\gamma^2 b)$, where $r_\mu = 1.35 \times 10^{-17}$ m is the electromagnetic radius of the muon, $N_{bunch}$ are the number of muons, $\varepsilon_o$ is the 95% normalized transverse emittance and $b$ is the bunching factor, defined as the ratio of the average beam current to the peak current. In the present conditions only integer resonances are truly harmful and therefore $\Delta Q \le 1$. For the first "wide band cooler" at $p_\mu \approx$ 230 MeV/c we assume the relatively small $b \approx 1/15$ with $\varepsilon_o = \sqrt{6}\,\varepsilon_N/\beta\gamma$ and therefore $N_{bunch} \approx$ 3.3 x $10^{13}$ $\mu^\pm$/bunch, to be compared with the nominal value of 2 x$10^{13}$ $\mu^\pm$/bunch at the entrance of the cooler. Traversing the "wide band" cooler the surviving bunch intensities are reduced by about a factor 2 due to losses and decays. For the subsequent Guggenheim helix [18] a bunch factor $b \approx 1/9$ is assumed (i.e. 1/3 of the circumference), since the momentum is lower and there are no injection and extraction kickers. Consequently the corresponding Laslett limit is 1.23 x $10^{13}$ $\mu^\pm$/bunch for $\varepsilon_N/\beta\gamma$ of Figure 4, compared with the required value of 1.0 x $10^{13}$ $\mu^\pm$/bunch.

## 5.— Conventional facilities.

The main innovative component here described in detail is the practical and experimental realization of a full scale cooling demonstrator, a relatively modest and low cost system but capable to conclusively demonstrate "ionization cooling" at the level required for a Higgs factory and eventually as premise for a subsequent multi-TeV collider and/or a long distance n factory.

The additional, much more expensive but relatively conventional facilities necessary to realize the facility with the appropriate luminosity should be constructed only after the success of this "initial cooling experiment" has been conclusively demonstrated. A detailed design of these further phases is beyond the scope of the present paper, and only indicative choices are given.

In order to arrive at the chosen energy of 125 GeV, an acceleration system is progressively rising the energy of captured muons to $m_{Ho}/2$ with the help of recirculating linear accelerators (LRA). For instance with a eight turn arrangement the energy increase of the LRA is 7.8 GeV/pass, corresponding to a length of ≈500 m for an average gradient of 15 MeV/m.

Adiabatic longitudinal Liouvillian damping from $p_i$ = 100 MeV/c to $p_f$ = 62.5 GeV/c is increasing the final momentum spread $\Delta p_f$ from 0.7 MeV/c to 2.0 MeV/c and reducing the bunch length $L_{b,f}$ = ± 2.4 cm. At this momentum spread the resulting rate of the SM Higgs is about ½ of the value for zero accelerator width.

The circular beam collider is a relatively small SC ring with a typical radius of ≈ 50 m and two low β sections with a free length of about 10 m, where the two detectors are located. Following the design of Ankenbrandt et al. [21] at the crossing points typical values could be $\beta_x = \beta_y = \beta^* = 4$ cm. The bunch transverse r.m.s. size is ≈ 0.2 mm and the µ–µ tune shift is 0.086. A luminosity of 0.6 x $10^{32}$ cm$^{-2}$ s$^{-1}$ is achieved with 6.1 x$10^{12}$ µ/bunch. The SM Higgs rate is then ≈ 6600 ev/year in each of the two detectors. The value β* = 4 cm is not a critical parameter and values of β* as small as ≈ 1.0 cm may be possible, with the corresponding luminosity increases, but only provided that the longitudinal dimension of the bunch is of comparable size, which requires an appropriate longitudinal emmittance. The expected beam-beam tune-shift is the reasonable value of 0.071

An important background which has to be carefully analyzed is coming from the ≈ 6 x $10^{12}$ muons /bunch from $\mu \rightarrow e \nu_\mu \nu_e$ decays, emitting 5 x $10^{11}$ $e^\pm/s/meter$ in a narrow average cone of $\langle \theta \rangle$ ≈ 1.6 mr from the beam axis, producing an average electron showers power of 1.6 kW/m. This amount of power, due to electron showering is comparable to the one in the case of synchrotron radiation from TLEP [5] or SuperTRISTAN [6]) and it must be handled with an appropriate geometry of the locations of the shower stoppers.

In addition in specific locations the angular distribution of the decay electrons and positrons should be used in order to measure and control as a function of time the individual muon bunches observing the g-2 precession frequency of polarized decays.

The present uncertainty with which the mass of the H$^o$ has been so far measured by ATLAS[1] and CMS [2] has been of the order of about 1 GeV. No doubt additional measurements will improve this result and a window of the order of ± 200 MeV should become reasonable in the forthcoming future. The first step of a future $\mu^+ - \mu^-$ collider should therefore be the one of finding the position of its signal in view of its actual very narrow resonance width. This is particularly favourable for the (WW) channel that has a substantial branching ratio of 22% according to the SM. The competing non-resonant background signal in the vicinity of the Ho is extremely small, of the order of ≈ 7 fb. In the bin closest to the signal for a Gaussian convoluted beam spread of 7 MeV (Figure 1) and for an energy scan in steps of 10 MeV (40 steps) the presence of the Higgs will record a minimum H$^o$ signal of 750 fb. An initial exposure of 10 effective days at 3 x $10^{31}$ cm$^2$ luminosity corresponds to 2.6 x $10^{37}$ cm$^2$ and > 20 events for the maximum of the (WW) resonant signal with a negligible background in each of the two experimental detectors.

Once the actual location of the H$^o$ resonance has been identified with the $\mu^+ - \mu^-$ collider, the actual value of its mass can be determined with a remarkable pre-cision observing the g-2 precession frequency of polarized muon decays. Raia and Tollestrup [20] have investigated the feasibility of determining the precise energy of the circulating muon beams (once accelerated to the final energy of 62.5 GeV) with the turn by turn variation of the electron energy spectrum produced by the decay $\mu \rightarrow e\nu\nu$. This variation based on a non-zero value of (g – 2) for the muon and of a finite polarization of the beam. This angular frequency/turn in the energy spectrum of the decay electrons $\omega = 2\pi\gamma(g-2)/2 \approx -0.7 \times 2\pi$ can be measured with high precision. Because of the narrow width of the Higgs boson, it is mandatory control the energy of the individual muon bunches to a precision of a few parts in a million. With adequate statistics it may then be feasible to determine the mass of the $H^o$ particle to the order of 100 keV, i.e. $\delta E/E \approx 10^{-6}$.

### 6.— Conclusions.

The $\mu^+ - \mu^-$ collider for the Higgs Factory is a relatively small circular high energy lepton collider that can be situated for instance within the CERN or FNAL sites. Their main parameters are given in Table 1 and Table 2. However it requires two major developments, namely:

1) the production and collection of an excess of $10^{12}$ $\mu^\pm$/bunch and
2) A 6D phase compression to a specified amount $\varepsilon_{6D} \approx 10^{-6}$

The initial experimental realization of a full scale cooling demonstrator represents its main innovative component, in the form of a relatively modest and low cost system but capable to conclusively demonstrate "ionization cooling" at the level required for a Higgs factory and eventually as premise for a subsequent multi-TeV collider and/or a long distance ν factory.

The full scale demonstrator can be initially explored with the help of very low intensity beam of muons which is already available in a number of different accelerators. All other conventional elements necessary to realize the facility with the appropriate luminosity, namely (1) the high intensity proton accelerator, (2) the pion/muon production target, (3) the subsequent muon acceleration and (4) the accumulation in a storage ring may be constructed only after the success of this initial cooling experiment has been conclusively demonstrated.

### 7.— Acknowledgments.

The author would like to acknowledge the Istituto Nazionale di Fisica Nucleare (INFN) of Italy and in particular his president prof. Nando Ferroni for interest and support. A similar paper entitled *"The Case for a Muon Collider Higgs Factory"* by Y. Alexahin et al. has just been published as arXiv:1307.6129.

Table 1. Tentative parameters of the complete demonstrator

| | | |
|---|---|---|
| *Proton beam* | | |
| Beam power | 5.0 | MWatt |
| Energy | 6-25 | GeV |
| Repetition rate | 15 | Sec-1 |
| *Muon collection* | | |
| Solenoidal field | 20 | Tesla |
| Momentum | 190-260 | MeV/c |
| Muons (each sign) | $2 \times 10^{13}$ | µ/bunch |
| Lifetime at 220 MeV/c | 5.061 | µs |
| Decay path | 1368 | m |
| *Wide band cooler* | | |
| Type of accumulator | RFODO | |
| Focussing | solenoids | |
| RF frequency | 200 | Mc/s |
| Ioniz. Coolant, wedge shaped | LiH | |
| Kickers | 2 | Inj.,Extr. |
| *Muon decelerator* | | |
| Input momentum | 220 | MeV/c |
| Exiting momentum | 100 | MeV/c |
| $LH_2$ wedge absorber (aver.) | 2.72 | m |
| *Guggenheim helix* | | |
| Average momentum | 100 | MeV/c |
| Type of accumulator | FODO | |
| Focussing | quadrupoles | |
| RF frequencies | 400, 800 | Mc/s |
| Ioniz. Coolant, wedge shaped | $LH_2$ | |
| Transv. normal emittance, final | 0.70 | mm rad |
| Long. normal emittance, final | 0.41 | mm rad |
| Rms energy spread, final | 740 | keV |
| Half bunch rms length, final | 4.3 | cm |

Table 2 Main parameters of the Higgs Factory

| | | |
|---|---|---|
| *Collider ring* | | |
| Circumference | 350.0 | m |
| Nominal energy at $H_o$ peak | 125 | GeV |
| Nominal muon momentum | 62.50 | GeV/c |
| Muons/bunch (each sign) | $6 \times 10^{12}$ | µ/bunch |
| Final lifetime: | 1.295 | ms |
| Mu decay length: | 388.6 | km |
| Average number of turns: | 1110. | |
| No effective luminosity turns: | 555.2 | |
| Crossings/sec: (at 15 hz) | 8328. | |
| Beta value at crossing point | 4.0 | cm |
| *Indicative performance* | | |
| $H_o$ peak cross section | $2.00 \times 10^{-35}$ | cm2 |
| Luminosity | $0.63 \times 10^{32}$ | cm-2 s-1 |
| $H_o$ events/y ($10^7$ s)/ each cross: | 12500 (*) | |
| $H_o$ reduction due to finite $\Delta E/E$ | 0.5 | |
| Bunch transv. rms size | 197.5 | microns |
| Beam-beam tune shift | 0.071 | |
| Final bunch half-length | 2.4 | cm |
| Final $\Delta p$ muon | 2.0 | MeV/c |
| Final $\Delta p/p$ muon | $3.2 \times 10^{-5}$ | |
| rms $\Delta E/E$ at $H_o$ resonance | $2.4 \times 10^{-5}$ | |

**8.—   References.**

[1]     ATLAS Collaboration, Phys. Lett. B 716 (2012) 1-29

[2]     CMS Collaboration, Phys. Lett. B 716 (2012) 30-61

[3]     J. Wess, B. Zumino, Nuclear Physics B 70 (1974) 39.

J. L. Gervais, B. Sakita, Nuclear Physics B 34 (1971) 632.

D.V. Volkov, V.P. Akulov, Pisma Zh.Eksp.Teor.Fiz. 16 (1972) 621; Phys.Lett. B46 (1973) 109; V.P. Akulov, D.V. Volkov, Teor.Mat.Fiz. 18 (1974) 39

[4]     R.S. Gupta, H. Rzehak, J.D. Wells, *"How well do we need to measure Higgs boson couplings?"*, arXiv:1206.3560 (2012).

J. Elias-Miro et al *"Higgs mass implications on the stability of the electroweak vacuum"* arXiv. 1112.3022

[5]     M. Koratzinos et al. *"TLEP: A High-Performance Circular e+e- Collider to Study the Higgs Boson"*, arXiv:1305.6498 [physics.acc-ph],2013

[6]     K. Oide, SuperTRISTAN , KEK meeting 13 Febuary 2012

[7]     D. B. Cline, "Physics potential of a few 100-GeV mu+ mu- collider," Nucl. Instrum. Meth. A 350, 24 (1994); V Barger et al., Physics Reports 286 (1997) 1-51;

[8]     G. I. Budker, in Proceedings of the 7th International Conference on High Energy Accelerators, Yerevan, 1969 (Academy of Sciences of Armenia, Yerevan, 1970), p. 33;. N. Skrinskii and V.V. Parkhomchuk, Sov. J. Part. Nucl., 12, 223 (1981); V.V. Parkhomchuk and A.N. Skrinsky, *Ionization cooling: physics and applications*, Proc. 12th Int. Conf. High Energy Phys., 1983, p. 485; E. A. Perevedentsev and A. N. Skrinsky, Proc., 12th Int. Conf. on High. Energy Accelerators, p. 481 (1993); J. Gallardo, R. Palmer, A. Tollestrup, A. Sessler, A. Skrinsky et al., "$\mu+$ $\mu-$ *Collider: A Feasibility Study*", DPF/DPB Summer Study on New Directions for High Energy Physics, Snowmass, Colorado, 25 Jun – 12 Jul 1996; BNL-52503, Fermilab - Conf - 96 – 092, LBNL – 38946; A.N. Skrinsky, *"Ionization cooling and muon collider"*, Proc. Workshop on Beam Dynamics and Technology Issues for muon Colliders, J. Gallardo, Ed., AIP Conf. Proc. 372, 1996, p. 133.

[9]     D. Neuffer, Part. Accel., 14 , 75 (1984) 75; D.V. Neuffer , Nucl. Instrum. Meth A350, 27 (1994); D. Neuffer, Principles and applications of muon cooling, Proc. 12th Int. Conf. High Energy Phys., 1983, p. 481; D. Neuffer, *"Recent Results on Muon Capture for a Neutrino Factory and Muon Collider"* NMFCC Note 520 (January, 2008); D. Neuffer, arXiv: 1207.4056; Proceedings of the First Workshop on the Physics Potential and Development of $\mu + \mu-$ Colliders, Napa, California (1992; D. Neuffer, "Colliding Muon Beams at 90-GeV, FERMILAB-FN-0319, (1979); D. Neuffer, IEEE Trans. Nucl. Sci. 28, 2034 (1981);

[10]    Ch, M. Ankenbrandt and al. Phys.Rev.ST Accel. Beams 2 (1999) 081001 physics/9901022 BNL-65623, FERMILAB-PUB-98-179, LBNL-41935, LBL-41935; S. Ozaki, R. Palmer, M. S. Zisman, J. Gallardo, editors, *Feasibility Study-II of a Muon-Based Neutrino Source*, BNL-52623, June, 2001; John F. Gunion, "Physics at a muon collider," AIP Conf. Proc. 435 (1998) 37, http://arXiv.org/pdf/hep-ph/9802258 ; C. Quigg, *"Physics with*

*a millimole of muons,"* AIP Conf. Proc. 435 (1998) 242, http://arXiv.org/pdf/hep-ph/9803326

[11]  M. M. Alsharo'a et al., *"Recent progress in neutrino factory and muon collider research within the Muon Collaboration,"* Phys. Rev. ST Accel. Beams 6 (2003) 081001: C. Ankenbrandt et al. *"Muon Collider Task Force Report,"* Fermilab -TM -2399 -APC (Jan 2008), S. Geer, *"Muon colliders and neutrino factories",* Ann. Rev. Nucl. Part. Sci. 59 (2009) 347

[12]  MICE Collaboration, *"An International Muon Ionization Cooling Experiment"*, proposal to RAL, MICE note 21, January 2003.

[13]  see for instance: C. Rubbia, Landau Nobel Meeting 2011; D. Neuffer, 14th International Workshop on Neutrino Factories, Super Beams and Beta Beams (NuFact2012), July 2012, Willianinsburg, Va, USA; C. Rubbia, XV International Workshop on Neutrino Telescopes (Venice, Italy) - March 11-15, 2013; Y. Alexahin et al., *"The Case for a Muon Collider Higgs Factory"* arXiv:1307.6129.

[13]  M. Reiser, *"Theory and Design of Charged Particle Beams",* (John Wiley & Sons Inc, New York, 1994); J.D. Jackson, *"Classical Electrodynamics"*, third edition, (John Wiley & Sons Inc., New York, 1998).

[14]  H.G. Kirk et al., *"A 15-T Pulsed Solenoid for a High- power Target Experiment",* Proc. 2008 Eur. Part. Accel. Conf. (Genoa, Italy, July 2008), WEPP170.

[15]  K.T.McDonald, Princeton Report/µµ/98-17 (1998), updated Feb. 2000

[16]  K. W; Robinson, Phys. Rev. 111,373 (1958)

[17]  V. Balbekov, *"Investigation of RFOFO-like Cooling Rings"* MUC-NOTE-THEORY-263, 2002. [14]; V. Balbekov, *"Simulation of RFOFO Ring Cooler with Tilted Solenoids,"* MUC-NOTE-THEORY-264, 2002; V. Balbekov, in Proceedings of the Particle Accelerator Conference, Portland, OR, 2003 (Ref. [6]), http:// accelconf.web.cern.ch/Accel/Conf/ p03/PAPERS/ WPAE033.pdf.; V. Balbekov, S. Geer, N. Mokhov, R. Raja, and Z. Usubov, in Proceedings of the 2001 Particle Accelerator Conference, Chicago, IL, 2001 (IEEE, Piscataway, NJ, 2001), http://accelconf.web.cern.ch/Accel/Conf/p01/ PAPERS/FPAH081.pdf. ; V. Balbekov and A. van Ginneken, in Physics Potential and Development of mu-mu Colliders, edited by D. B. Cline, AIP Conf. Proc. No. 441 (AIP, New York, 1998), p. 310. [16]

[18]  R. Palmer et al., Phys. Rev. ST Accel. Beams 8, 061003 (2005); R. Palmer et al. *"The RFOFO Ionization Cooling Ring for Muons"* arXiv:physics /0504098 v1 14 Apr 2005; J.S. Berg et al. , AIP Conf. Proc. 721, 391 (2004); R.C. Fernow et al, BNL-71409-2003 and 2003 Particle Accelerator Conference (PAC2003), Portland, Oregon, US; R. Palmer, J. Phys. G: Nucl. Part. Phys. 29 (2003) 1577–1583; J. Berg et al. Phys.Rev.ST Accel.Beams 9, 011001 (2006)

[19]  A. Garren et al., Nucl. Instrum. & Meth. A 654 (2011) 40–44

[20]  R. Raja and A. Tollestrup, Phys Rev. D **58**, 013005 (1998)

[21]  C.M. Ackenbrandt et al., Phys.Rev. ST Accel. Beams, 2,08001 (1999)

9.— Figures

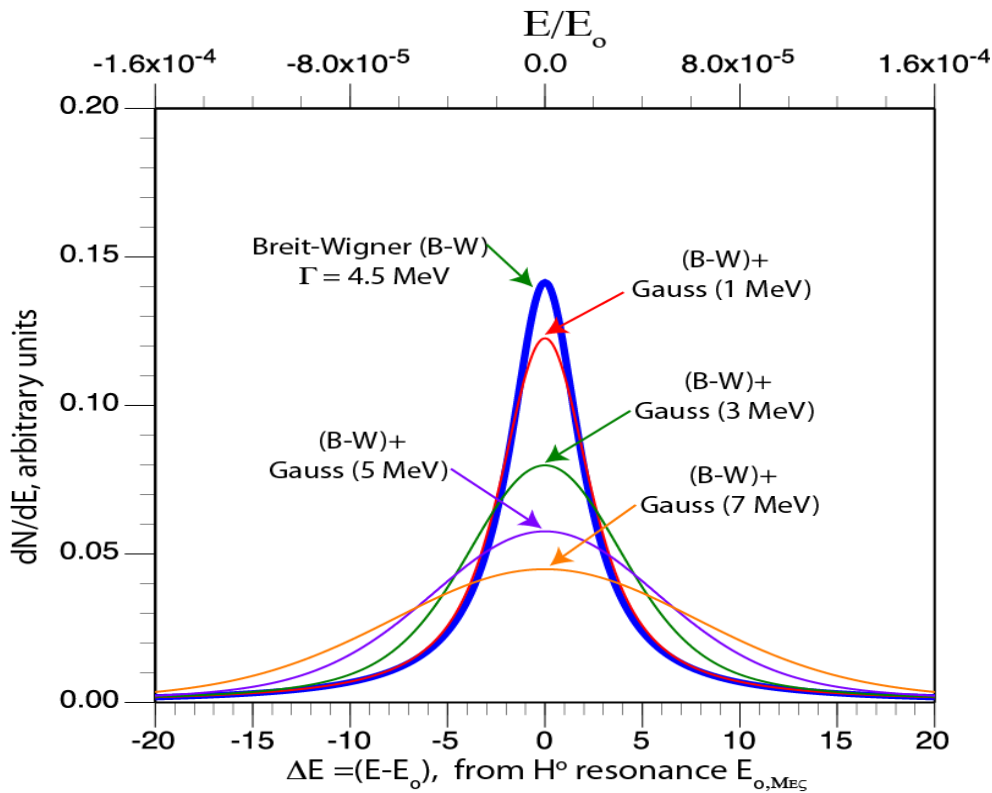

**Figure 1.** Breit-Wigner H° line width according to the Standard Mode(SM). The line has been convoluted with a Gaussian r.m.s. width coming from the beam resultant width. Values for 1, 3, 5, and 7 MeV are shown.

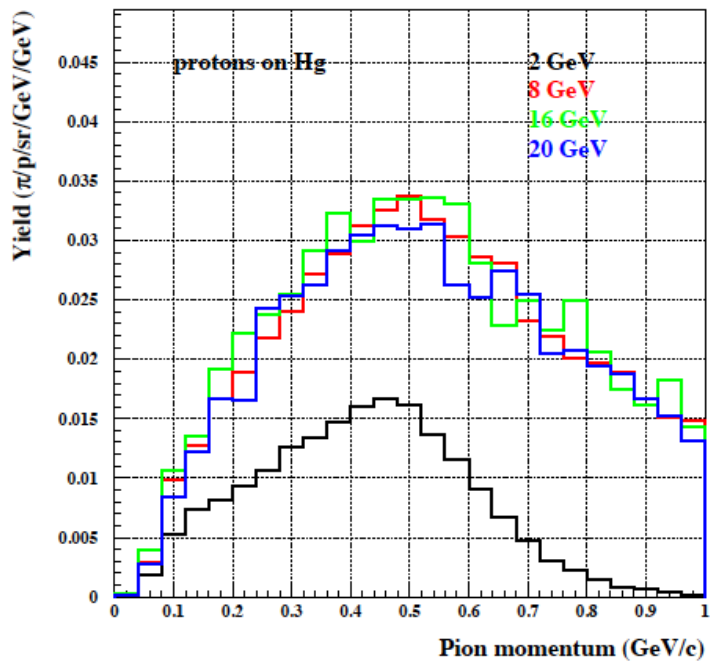

**Figure 2.** Pion yield as a function of the pion momentum as a function of a given proton power of the beam (the number of protons is inversely proportional to the beam energy). Note the remarkable similarities of the pion yields for proton energies > 8 GeV.

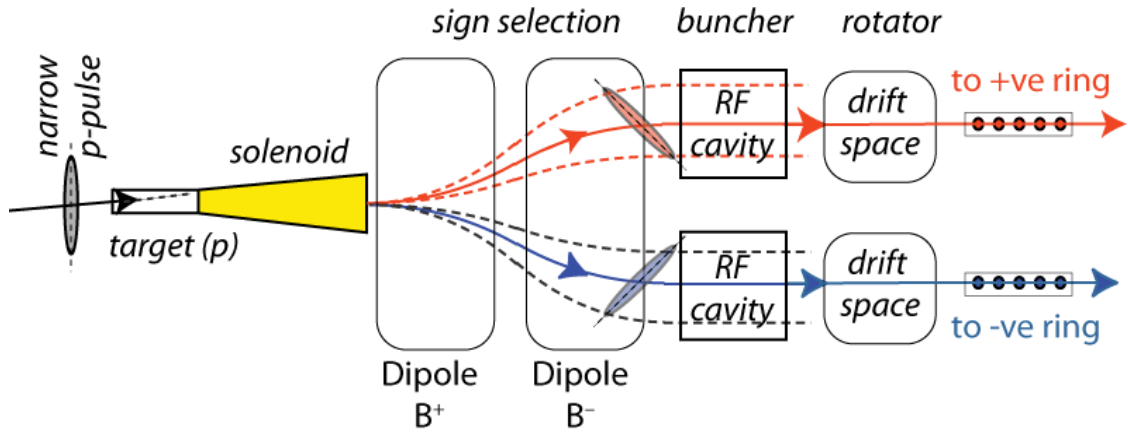

**Figure 3.** Short, intense bunches of protons are hitting a target to produce pions, that then decay into muons. Target is immersed in the high field (20 Tesla) solenoid and beams of both signs are focussed by the tapered solenoid. Positive and negative particles are separated with a pair of transverse dipoles. It may be convenient to decrease the momentum spread with a corresponding increase of the beam longitudinal emittance for instance with the help of a RF buncher followed by a drift space. Other options are also described in the text.

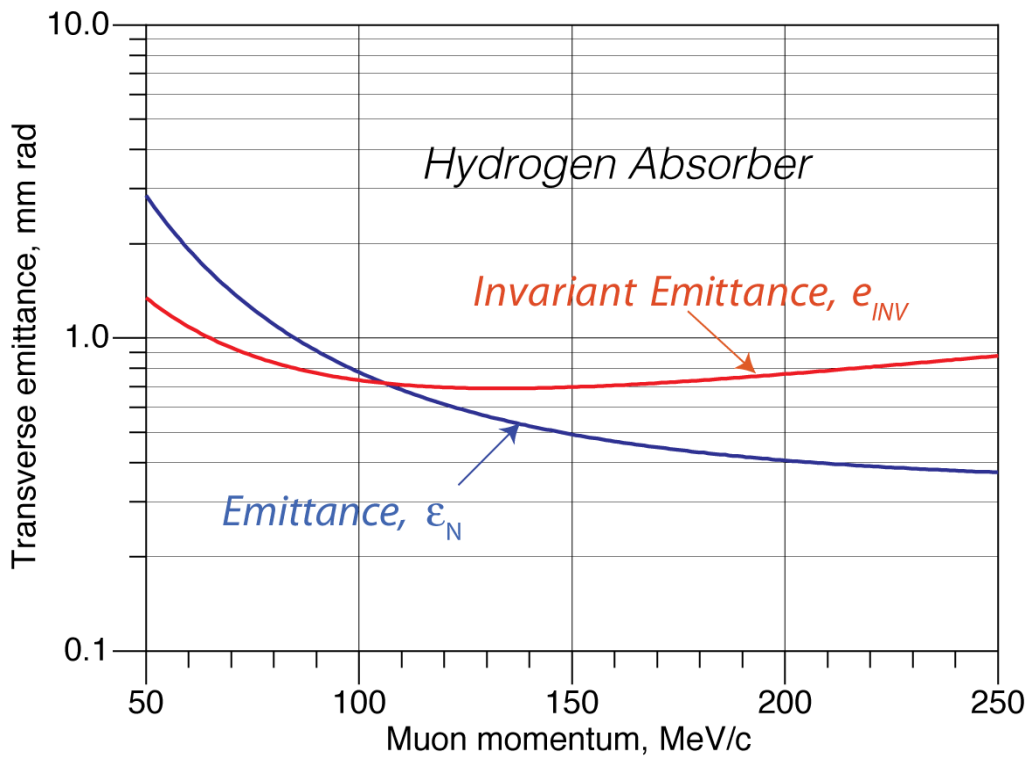

**Figure 4.** The equilibrium emittance $\varepsilon_N$ and its invariant $\varepsilon_N/\beta\gamma$ are shown as a function of the stored muon momentum for the case of a liquid hydrogen absorber. The invariant emittance is remarkably constant over the full range of optimal muon momenta.

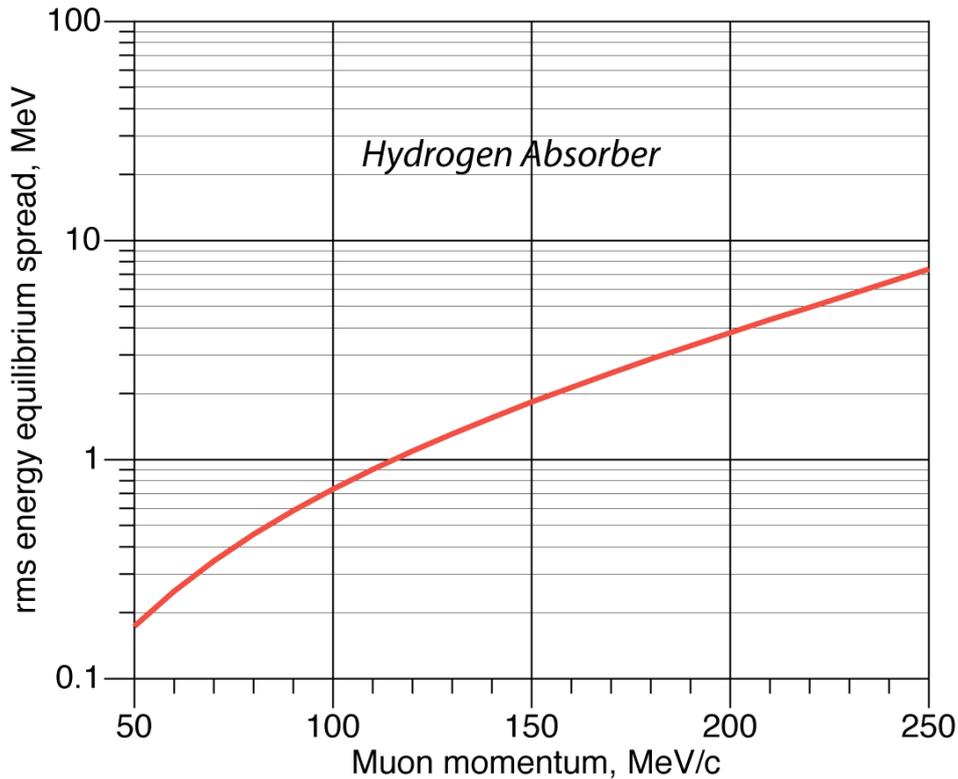

**Figure 5,-** R.m.s. equilibrium energy spread as a function of the muon momentum $p_\mu$ for the case of liquid hydrogen absorber. The energy spread is a very quickly decreasing function of the muon momentum. The energy spread is adequate for the requirements of a SM Higgs only for $p_\mu \leq 100$ MeV/c.

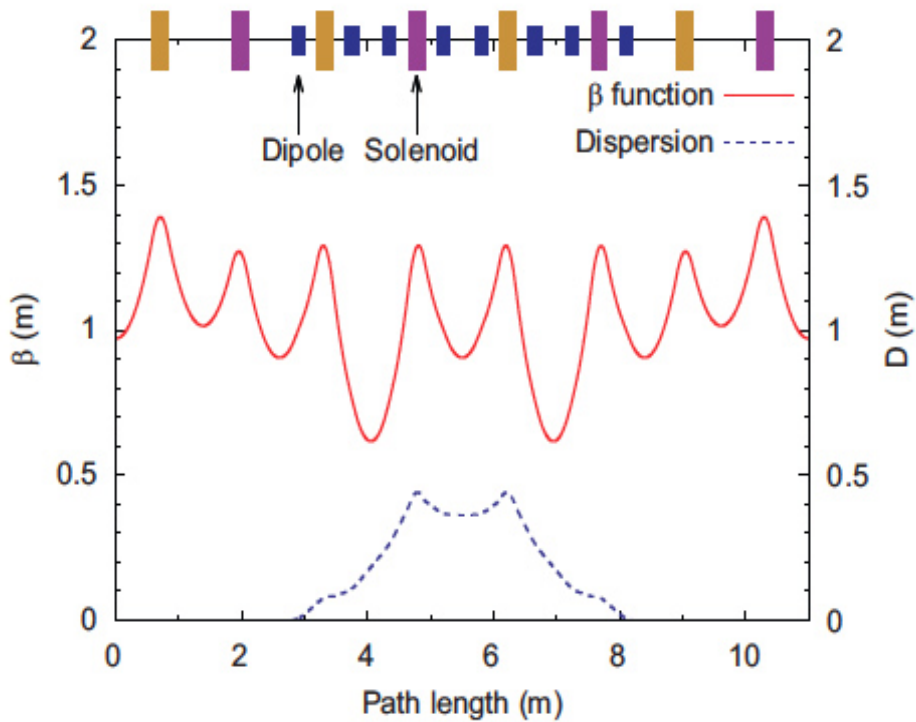

**Figure 6**. A tentative design [] of the "wide band beam". The four sided ring has 90° arcs, each with 8 dipoles separated by solenoids. Arcs are achromatic horizontally and vertically. Dispersion D is zero in the straight sections between the arcs. An elaborate kicker in the straight section and a superconducting flux pipe for the injected beam are used.

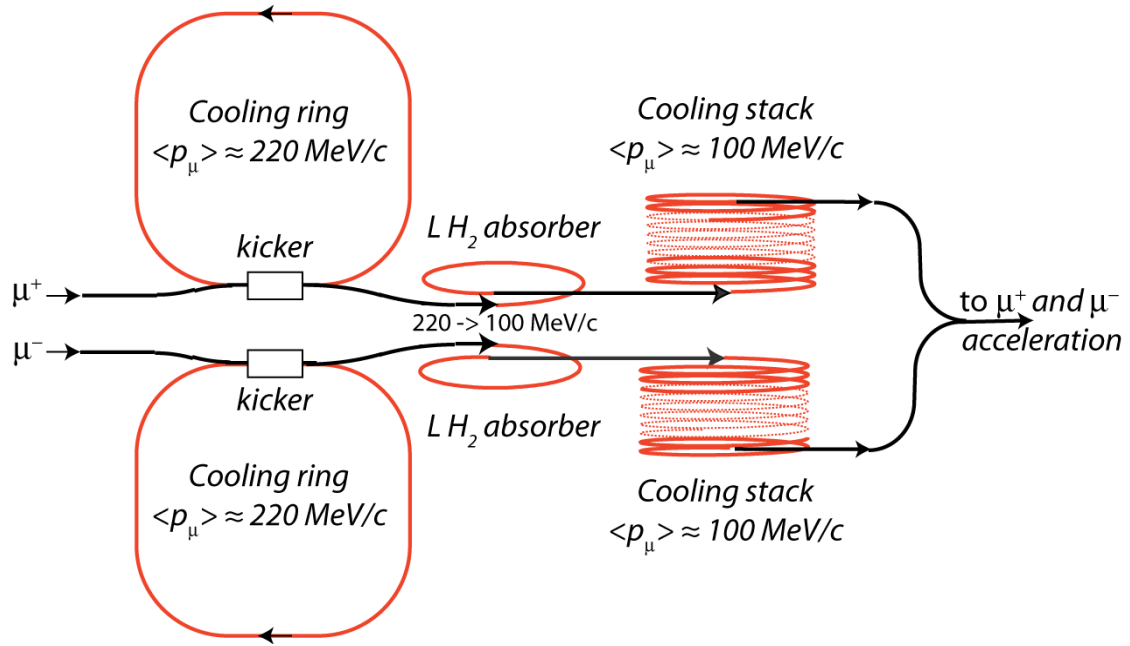

**Figure 7.** General layout of the fullscale demonstrator. Muons of both signs are injected in the "wide band" cooling ring over a momentum byte of ±20% centred around $p_\mu$ = 220 MeV/c. After cooling has approached the equilibrium value, the beam is extracted and its momentum reduced to about 100 MeV/c a corresponding to a kinetic energy $T_\mu$ = 39.7 MeV with the help of a wedged liquid hydrogen absorber about 270 cm long. A low $\beta^*$ channel is required in order to reduce the blowup of the beam due to multiple scattering and straggling. The beams are then each injected in a Guggheneim cooling helix in order to cool the longitudinal momentum spread to the required r.m.s. value of $\Delta E \approx 0.7$ MeV/c.